\documentclass[12pt]{article}
\usepackage{graphicx}
\usepackage[dvips]{color}

\begin{document}

\author{M. I. Jaghoub$^1$\thanks{
E-mail address: mjaghoub@ju.edu.jo} \hspace{0.3ex}, \ G. H. Rawitscher$^2$ \\
{\small {\ $^1$ Department of Physics, University of Jordan, P. C. 11942,
Amman, Jordan}}\\
{\small {$^2$ Department of Physics, University of Connecticut, Storrs, CT
06269, USA}}}
\title{Evidence of nonlocality due to a gradient term in the optical model}
\maketitle

\begin{abstract}
We demonstrate that the presence of a velocity-dependent term in the phenomenological
optical
potential simulates a source of nonlocality. This is achieved by showing
that, in the interior of the nucleus, the nonlocal wave functions are
different from the corresponding local ones obtained in the absence of the
velocity-dependent term in accordance with the Perey effect. It is also
shown that the enhancement or suppression of the nonlocal wave function is
energy as well as angular momentum dependent. The latter is in line with the
results of previous works that introduced parity dependent terms in the
conventional optical potential.
\end{abstract}


\newpage

\section{Introduction}

It is a well known fact that the nonrelativistic nucleon-nucleus optical
potential is nonlocal and energy dependent \cite%
{Feschbach_Ann,Feschbach_book}. Nevertheless, a common approach to studying
the elastic $NA$ scattering is the use of phenomenological local optical
models, and a variation of the model parameters with the incident projectile
energy can be taken as a sign of the presence of nonlocal effects in the
scattering process \cite{Fraser}. For example, a recent work presented local
and global optical potentials for neutron and proton elastic scattering from
nuclei that fall in the mass range $24\geq A\geq 209$ \cite{Koning}, while
nucleon incident energies ranged from 1 keV to 200 MeV. The authors obtained
excellent elastic angular distribution fits using local optical potentials
each consisting of 20 fit parameters, however, the strength of the central
real part showed the largest variation with incident energy. Another measure
of nonlocality is given by the size of the Perey damping factor, as will be
discussed in the present study for the case of the presence of a gradient
term in the optical potential. Such a gradient term, also denoted as
velocity dependent term, is usually absent in conventional optical model
potentials, but has been introduced in a previous study \cite{MJ-GR-PRC},
and will be the object of further analysis in the present investigation.

The nonrelativistic optical potential contains several sources of
nonlocalities. One such nonlocality is due to the Pauli exclusion principle
and can be taken into account by antisymmeterizing the wave function as in
the Hartree-Fock theory \cite{Austern}. Further, in the Hartree-Fock
equation, the nonlocality due to exchange effects can be expressed in terms
of a spatially variable mass \cite{Ring}. More recently, the importance of
accounting for the Pauli principle at incident energies greater than 25 MeV
has been established \cite{Amos-2000NP}. At low energies the importance of
the Pauli exclusion principle has been investigated in collective-model
coupled-channel calculations \cite{Canton-05PRL}. A second source of
nonlocality is due to the coupling of the inelastic excitations to the
nuclear ground state during the scattering process \cite{Feschbach_book},
that gives rise to a nonlocal dynamic polarization potential in the elastic
channel. Since this nonlocality is difficult to take into account
rigorously, it is usually accounted for by numerically coupling a few
important inelastic channels to the elastic channel \cite{Geramb,RAW_NP}.
However, when the energy of the incident nucleon is low enough, the
scattering process is strongly affected by the individual discrete states of
the target nucleus. A recent study that employed the multichannel algebraic
scattering theory (MCAS) to explore the dynamic polarization potential \cite%
{Fraser,Canton} for low incident energies also showed that the resulting
optical potential was strongly nonlocal and energy dependent.

Recently, we introduced a novel gradient term in the Schr\"{o}dinger
equation describing the $NA$ elastic scattering process. We interpreted this
term as a change in mass of the incident nucleon when moving inside the
nuclear matter of the target nucleus \cite{MJ-GR-PRC}. The mass change could
be a consequence of the nucleon-nucleon interaction being affected by the
presence of surrounding nucleons, like a surface tension which is present
only at the surface of the target nucleus. This view is supported by the
fact that our model, in Ref.\  \cite{MJ-GR-PRC}, resulted in excellent fits
to the $N-^{12}C$ elastic scattering data when the coefficient of the
derivative term was assumed to be proportional to the gradient of the
nuclear density, which is most important at the nuclear surface.
Furthermore, our model reproduced well the large angle scattering minima
that are usually associated with nonlocalities \cite{Rawitscher-L-69,Clark}.
However, it is still not yet clear whether the gradient term also simulates
other sources of nonlocal effects in the phenomenological optical model
like, for examples, the Pauli exclusion principle and coupling of the
elastic channel to the inelastic ones.

We formulated the mass dependence by introducing a Kisslinger type potential
into the conventional optical model \cite{Kisslinger}. This is achieved by
writing the Schr\"{o}dinger operator as 
\begin{equation}
-\mathbf{\nabla }\cdot \frac{\hbar ^{2}}{2m^{\ast }(r)}\mathbf{\nabla +}V-E=-%
\frac{\hbar ^{2}}{2m_{0}}\mathbf{\nabla }^{2}+\hat{V}(r,p)-E,
\end{equation}

where the resulting potential $\hat{V}$ is velocity-dependent and has the
form: 
\begin{eqnarray}
\hat{V}(r,p) &=&V(r)+\frac{\hbar ^{2}}{2m_{0}}\mathbf{\nabla }\cdot \rho (r)%
\mathbf{\nabla }  \nonumber \\
&=&V(r)+\frac{\hbar ^{2}}{2m_{0}}\left[ \rho (r)\nabla ^{2}+\mathbf{\nabla }%
\rho (r)\cdot \mathbf{\nabla }\right] .  \label{vdp}
\end{eqnarray}%
The velocity-dependent potential $\hat{V}(r,p)$ plays the role of a nonlocal
potential $V(r,r^{\prime })$. In fact there is an equivalence between a
velocity-dependent (or momentum-dependent) potential and a nonlocal
potential $V(r,r^{\prime })$ \cite{Feschbach_Ann}. This can be seen
by interpreting the gradient term that acts on the wave function as the
first term of a Taylor series that displaces the wave function from point $%
\vec{r}$ to a different location. Furthermore, the spatially
variable effective mass $m^{\ast }(r)$ is assumed to have the 
\begin{equation}
\frac{1}{m^{\ast }(r)}=\frac{1}{m_{0}}(1-\rho (r))  \label{mass}
\end{equation}%
where $\rho (r)$ is an isotropic function of the radial variable $r$ that
expresses the change of the projectile's mass with distance $r$ from the
center of the target nucleus \cite{MJ-GR-PRC}. The notion of an effective
mass has also been introduced in relation to relativistic optical models
that give rise to significantly reduced effective masses \cite%
{Serot}. For example, the relativistic effects on the nucleon mean free path
in the intermediate energy range were investigated in Ref.\  \cite{Cheon}. In
addition, the idea of an effective mass was considered for the specification
of a microscopic optical model potential Ref.\  \cite{Amos-2000NP}. At this
point we stress that the gradient term in our model is different from the
Darwin term present in the nonrelativistic equivalent of the
Dirac-based relativistic formulation  \cite{Arnold} because, firstly, our
effective mass is real while the Darwin term is complex. Secondly, 
in the nonrelativistic equivalent of the Dirac formulation, the Darwin
term is closely coupled to the spin orbit potential while our gradient term
is independent of the spin orbit term \cite{MJ-GR-PRC}. Furthermore, unlike
the case for the relativistic models mentioned above, our gradient term is
expressed as an effective mass that has a radial dependence, rather
than in terms of an additional scalar potential that is added to the mass.
Finally, the effective mass formulation has also been introduced in the
field of condensed matter physics calculations \cite{Geller,Serra,Gomez}.

Other phenomenological forms of nonlocalities have also been considered such
as a parity-dependent term in the optical potential \cite{Mackintosh, Cooper}
for low energy scattering of protons from $^{16}O$. This parity dependence
was later justified tentatively in terms of the Feshbach channel coupling
effect \cite{Rawitscher-L-69,RG_94}. A presently much used nonlocality, 
first introduced by Frahn and Lemmer \cite{Frahn} and 
further developed  by Perey and Buck in the 1960ies, is 
given in the form of an integral kernel $K(r,r\prime )$ that acts on the
wave function. Perey and Buck also obtained a local equivalent
potential in the corresponding Schr\"{o}dinger equation \cite{Perey}.

A non local potential $V(r,r')$ acts on the nonlocal wave function $\psi _{NL}$
in the form $\int V(r,r')  \psi _{NL}(r')
dr',$ while a local potential $V(r)$ acts as a factor $%
V(r)\  \psi _{L}(r)$, where $\psi _{L}(r)$ is a local wave function. Like the local wave function, $\psi _{NL}$ is a function of one radial coordinate only. In the nuclear interior, a local potential leads to
local wave functions which are different from the corresponding nonlocal
ones obtained using a nonlocal potential, even though both forms may lead to
the same asymptotic phase shift. This is known as the Perey effect \cite%
{Austern}, and may be expressed quantitatively as 
\begin{equation}
\psi _{NL}(r)=F(r)\psi _{L}(r),  \label{Fr}
\end{equation}%
where $\psi _{L}(r)$ and $\psi _{NL}(r)$ are the local and nonlocal wave
functions respectively, while the Perey damping factor is defined as $%
F(r)^{2}$ \cite{Mackellar}. In the nuclear interior, an attractive nonlocal
potential results in a reduction in the nonlocal wave function ($F(r)<1$),
while for a repulsive potential the situation is reversed leading to an
enhancement of $\psi _{NL}(r)$, that is, $F(r)>1$. As discussed by Austern 
\cite{Austern}, nonlocality arises when one singles out from the overall
wave function that part which describes the motion in the elastic channel.
Outside the nucleus $\psi _{NL}(r)$ is the full physical wave function of
the system. However, inside the nucleus part of the wave function of the
system is hidden in the invisible (inelastic) channels, which results in a
suppression of the nonlocal elastic wave function compared to the local one.
Conversely, for a repulsive nonlocal potential, the particles of the target
nucleus are driven away from the incident particle. However, conservation of
the flux demands that the incident particle carries more flux in order to
compensate for the part of the channel space which has been suppressed \cite%
{Austern}. A more recent work employed nonlocal potentials in three-body
direct nuclear reaction calculations and an important nonlocality effect in
transfer reactions was reported \cite{Detluva}.

In this work our aim is to show that the presence of a velocity-dependent
part in the conventional, phenomenological optical potential simulates a
source of nonlocality. This is achieved by showing that, in the interior of
the nucleus, the nonlocal wave functions are different from the
corresponding local ones obtained in the absence of a gradient term ($\rho
(r)=0$) in accordance with the Perey effect. We shall also demonstrate that
this effect is angular momentum dependent.

\section{Nonlocality due to the gradient term}

In Ref.\  \cite{MJ-GR-PRC}, we considered a novel velocity-dependent optical
potential of the form given in Eq.\ (\ref{vdp}) based on the change of mass
of the incident nucleon, and obtained excellent fits for the elastic $%
N-^{12}C$ angular distribution data in the low energy range of $12-20$ MeV.
Most notably the model reproduced well the prominent large-angle,
backscattering minima which depend sensitively on the incident energy, and
which are a sign of the presence of nonlocalities \cite%
{Rawitscher-L-69,Clark}. The local potential consisted of central and spin
orbit potentials and was taken to have the form: 
\begin{equation}
V(r)=-V_{0}f(r,x_{0})+4ia_{w}W\frac{df(r,x_{w})}{dr}+20\left( \frac{\hbar }{%
\mu c}\right) ^{2}(V_{so}+iW_{so})\frac{1}{r}\frac{df(r,x_{so})}{dr}\vec{%
\sigma}\cdot \vec{I},  \label{Vr}
\end{equation}%
where $x_{0}$ stands for $(r_{0},a_{0})$ and so on for the rest of the
terms. In addition, the function $f(r,r_{j},a_{j})$ has the following
Woods-Saxon form: 
\begin{equation}
f(r,r_{j},a_{j})=\frac{1}{1+\exp [(r-r_{j}A^{1/3})/a_{j}]},
\end{equation}%
with $A$ being the mass number of the target nucleus. In order to reduce the
number of fit parameters (12) we used the same set of parameters for the
real and imaginary parts of the spin-orbit term. In what follows the
nonlocal wave function $\psi _{NL}(r)$ is the solution of the Schr\"{o}%
dinger equation corresponding to the velocity-dependent potential $V(r,p)$
given in Eq.\  (\ref{vdp}). The potential parameters are adjusted to obtain a
best fit to the experimental $N-^{12}C$ elastic scattering angular
distribution. The local wave function $\psi _{L}(r)$, however, is the
solution of the Schr\"{o}dinger equation corresponding to the equivalent
local potential $V(r)$ whose parameters are adjusted to (i)
fit experimental data and (ii) produce the same phase shift values as those
obtained using the nonlocal potential $V(r,p)$. In the local equivalent
potential case, the effect of nonlocalities in the scattering process are
manifested as a variation of the potential parameters with energy. This
procedure is not unique, but is guided by the conventionally used method to
fit angular distributions. The local wave function obtained in this way is
not equal to the trivially equivalent local wave function, which is obtained
by starting the fit in the presence of a velocity dependent term, and then
mathematically transforming the $d\  \psi _{NL}(r)/dr$ away by a
renormalization of the wave function $(1-\rho (r))^{1/2}\psi _{NL}(r)$ \cite%
{Ericson}.

In Sec.\  \ref{RD}, we demonstrate that the nonlocal wave function $%
\psi_{NL}(r)$ is indeed different form the local wave function $\psi_L(r)$
in the nuclear interior, which supports our hypothesis that the gradient
term introduced into the conventional phenomenological optical potential
simulates a source of nonlocality. In addition, we find that the enhancement
or suppression of the nonlocal wave function relative to the local one to be
angular momentum dependent. This is in line with earlier works that modeled
the $NA$ elastic scattering by introducing a parity-dependent term $(-1)^{l}$
into the optical potential \cite{Cooper}.

\section{Results and Discussion \label{RD}}

As described above, the parameters of the equivalent local potential $V
(r)$ given in Eq.\ (\ref{Vr}) were adjusted until (i) a best fit for the
elastic $N-^{12}C$ angular distribution data was obtained and (ii) the
resulting phase shifts were almost identical to the ones obtained
corresponding to the nonlocal fits in Ref.\  \cite{MJ-GR-PRC}. The fits for
the $12,14$ and $16$ MeV incident nucleon energies are shown in Fig.\  \ref%
{AngularFits1}, where the experimental data are taken from Ref.\  \cite{ENDF}%
. The equivalent local potential fit parameters as a function of the
incident energy and orbital angular momentum are shown in Table I.

\begin{table}[th]
\begin{tabular}{lccccccccc}
\hline \hline
&  &  &  &  &  &  &  &  &  \\[0.01ex] 
$E_{\mathrm{lab}}$ & $V_{0}$ & $r_{0}$ & $a_{0}$ & $W$ & $r_{w}$ & $a_{w}$ & 
$V_{so} + i W_{so}$ & $r_{so}$ & $a_{so}$ \\ 
MeV & MeV & fm & fm & MeV & fm & fm & MeV & fm & fm \\ \hline
&  &  &  &  &  &  &  &  &  \\[0.01ex] 
$12$ & $44.0$ & $1.33$ & $0.48$ & $7.2$ & $1.30$ & $0.34$ & 17 + 4 i & 1.10
& 0.08 \\ 
$14$ & $50.0$ & $1.24$ & $0.56$ & $5.8$ & $1.22$ & $0.53$ & 12 + 0 i & 1.05
& 0.11 \\ 
$16$ & $43.0$ & $1.40$ & $0.37$ & $6.0$ & $1.55$ & $0.28$ & 26 + 8 i & 1.00
& 0.11 \\ 
$18$ & $47.5$ & $1.44$ & $0.50$ & $8.7$ & $1.10$ & $0.41$ & 20 + 15 i & 0.91
& 0.09 \\ 
$20$ & $41.0$ & $1.61$ & $0.51$ & $8.0$ & $1.36$ & $0.40$ & 20 + 5 i & 0.80
& 0.15 \\ \hline \hline
\end{tabular}%
\caption{Conventional optical model fit parameters for the $N-^{12}C$
elastic scattering angular distribution. The potential terms are given by
Eq.\ (\protect \ref{Vr}).}
\label{Table1}
\end{table}
Clearly, the overall behavior of the angular distributions has been
reproduced. However, the fine details of the differential cross sections are
less well described compared to the corresponding fits obtained using the
phenomenological velocity-dependent optical potential \cite{MJ-GR-PRC}. The
large angle back scattering minima (associated with nonlocalities) at 18 and
20 MeV were poorly reproduced by the local optical potential and hence are
not shown.

Evidently, the fit parameters are energy sensitive, which indicates the
presence of sources of nonlocalities. Most notable among them being the
variation in the strength of the central potential $V_{0}$. The
corresponding nonlocal (velocity-dependent) optical potential fit
parameters (given in Tables I and II of Ref.\  \cite{MJ-GR-PRC}) are less
energy dependent indicating that the introduced gradient term only accounts
for part of the nonlocalities present in the optical model. In a furure work, we intend to investigate other sources of nonlocalities in the phenomenological optical potential.

In this work we have also investigated the suppression and
enhancement effects of the nonlocal wave function $\psi _{NL}(r)$ as a
function of the incident nucleon energy in addition to the orbital angular
momentum. The Perey factor $F(r)$ defined by Eq.\ (\ref{Fr}) has been
determined by comparing the maxima of the real parts of $\psi _{NL}(r)$ and $%
\psi _{L}(r)$ in the nuclear interior. Since the gradient term simulating
the nonlocality is purely real, the imaginary part of $\psi _{NL}(r)$ is
almost identical to the corresponding local one. Further, the local wave
function has been normalized such that it coincides with the nonlocal one in
the asymptotic region. The damping (or enhancement) factor $F(r)^{2}$ as a
function of the incident energy as well as the orbital angular momentum is
shown in Table \ref{Table2}.

\begin{table}[ht]
\begin{tabular}{lcccccc}
\hline \hline
&  &  &  &  &  &  \\[0.2ex] 
$E_{\mathrm{lab}}$ & \multicolumn{6}{c}{angular momentum quantum number $l$}
\\ \cline{2-7}
&  &  &  &  &  &  \\ 
MeV & 0 & 1 & 2 & 3 & 4 & 5 \\ \hline
&  &  &  &  &  &  \\[0.1ex] 
$12$ & $1.746$ & $0.838$ & $0.876$ & $1.451$ & $1.00$ & $1.00$ \\ 
$14$ & $1.44$ & $0.779$ & $0.835$ & $1.477$ & $1.00$ & $1.00$ \\ 
$16$ & $1.134$ & $0.882$ & $1.171$ & $0.918$ & $1.00$ & $1.00$ \\ 
$18$ & $1.355$ & $0.636$ & $1.227$ & $2.001$ & $1.00$ & $1.00$ \\ 
$20$ & $0.928$ & $0.700$ & $0.988$ & $1.522$ & $0.895$ & $1.00$ \\ 
\hline \hline
\end{tabular}%
\caption{Perey factor $F(r)^2$ as defined in Eq.\ (\protect \ref{Fr}). The
values are calculated at the peaks of the nonlocal and local wave functions
in the nuclear interior.}
\label{Table2}
\end{table}

\begin{figure}[ht]
\includegraphics[height=160mm, width=86mm]{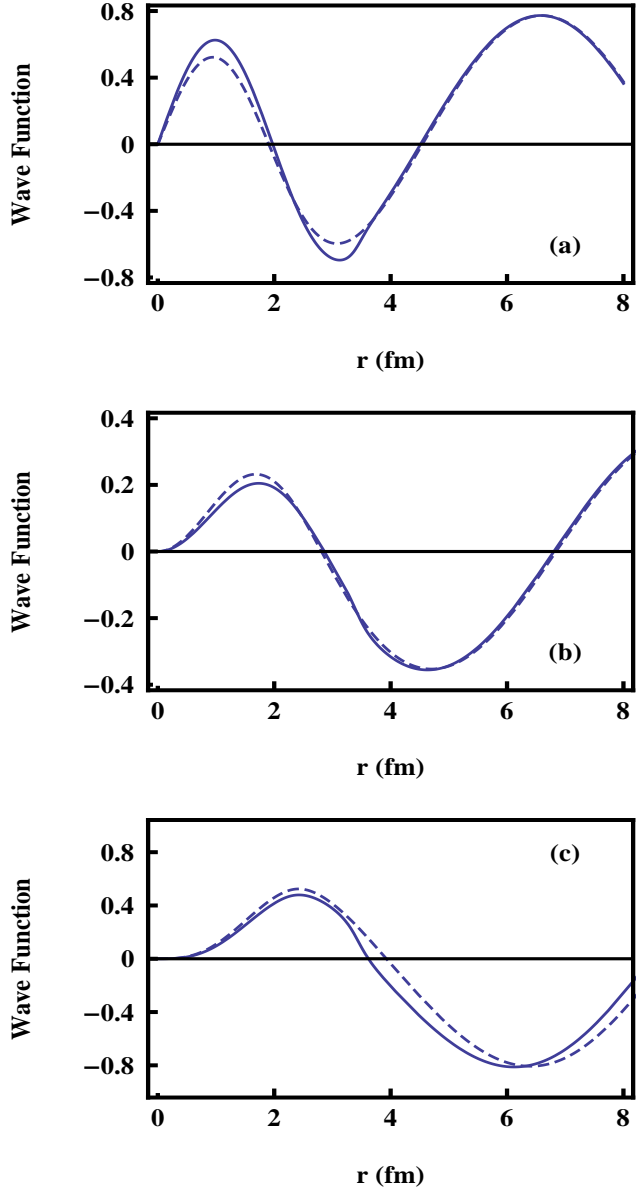}
\caption{ The real parts of the local (dotted) and nonlocal (solid) wave
functions corresponding to (a) $l=0$, (b) $l=1$ and (c) $l=2$ angular
momentum quantum numbers. The local and nonlocal wave functions correspond to the local and nonlocal potentials given by Eqs.\ (\ref{vdp}) and (\ref{Vr}). All figures correspond to $E=14$ MeV incident
neutron energy. }
\label{wfncs1}
\end{figure}

By inspecting Table \ref{Table2} it is evident that the suppression and
enhancement effects are both energy and angular momentum dependent. Apart
from the $20$ MeV incident energy case, where the angular distribution fits
are poorer than those obtained at lower energies, the $s$-wave nonlocal wave
function is enhanced. For $p$-wave scattering a suppression is noted for all
incident energies. Corresponding to $l=2$, however, we have a mixed
situation where the nonlocal wave function is suppressed for low energies
and enhanced corresponding to the higher $16$ and $18$ MeV ones. However,
the 20 MeV case stands out again and results in a small suppression. For the 
$l=3$ case, we have an enhancement at all energies apart from the 16 MeV
incident energy. For $l\geq 4$, the nonlocal wave function is too small in
the interior region of the nucleus to feel the effect of the gradient term,
which simulates the nonlocality, hence $F(r)^{2}$ reduces to close to unity.
As an illustration, in Figs.\  \ref{wfncs1} and \ref{wfncs2} we show the real
parts of the local $\psi _{L}(r)$ and nonlocal $\psi _{NL}(r)$ wave
functions corresponding to different angular momentum quantum numbers all at
the same $14$ MeV incident neutron energy. As shown in Fig.\  \ref{wfncs1}
(a), for the $s$-wave case the nonlocal wave function (solid line) is
clearly enhanced in the nuclear interior compared to the local one (dotted
line). Further, Figs.\  \ref{wfncs1} (b) and \ref{wfncs1} (c) corresponding
to $l=1$ and $l=2$ respectively show a suppression rather than an
enhancement. For $l=3$, the maximum of $\psi _{NL}(r)$ coincides with the
peak of the gradient term (dash-dotted line), which explains the distortion
of $\psi _{NL}(r)$ inside the nucleus as shown in Fig.\  \ref{wfncs2} (a).
Finally, for the $l=4$ case, the nonlocal wave function in the nuclear
interior is too small to be affected by the gradient term that simulates the
nonlocal effect and acts in the internal nuclear region. Therefore, the two
wave functions coincide in the interior region as can be seen in Fig.\  \ref%
{wfncs2} (b). The fact that $\psi _{L}(r)$ and $\psi _{NL}(r)$ coincide in
the external region (for all angular momenta) is guaranteed by the adopted
normalization method explained above. In addition, since the nonlocality is
simulated in terms of a purely real term only the real part of the nonlocal
wave function is affected, the imaginary part is almost identical with the
corresponding local one and hence is not shown in the figures.

\section{Trivially local equivalent potential \label{TLP}}

The Schr\"{o}dinger equation corresponding to a velocity-dependent potential 
$V(r,p)$ can be transformed into an equation for an equivalent local but
energy-dependent one, $U(r,E)$, for a trivially equivalent local
wave function $\chi _{L}(r)$ through the following transformation on the
nonlocal wave function $\psi _{NL}(r)$ \cite{Ericson};

\begin{equation}
\psi _{NL}(r)=\frac{\chi _{L}(r)}{\sqrt{1-\rho (r)}},  \label{trans}
\end{equation}%
which, as derived in \cite{jaghoubA28}, leads to the following
energy-dependent term in the equivalent local potential;

\begin{equation}
\frac{\rho (r)k^{2}}{1-\rho (r)},
\end{equation}%
where, as defined in Ref.\  \cite{MJ-GR-PRC}, $\rho $ is given by 
\begin{equation}
\rho (r)=\rho _{0}a_{\rho }\ d/dr\left \{ \frac{1}{1+\exp [(r-r_{\rho
}A^{1/3})/a_{\rho }]}\right \} ,  \label{rho}
\end{equation}
In this case the values of the fitting parameters for the nonlocal potential  $\hat{V}(r,p)$, given by Eq.\ (\ref{vdp}), and the local but energy dependent one $U(r,E)$ are identical. Consequently, $\psi_{NL}(r)$ and $\chi_L(r)$ share the same values of the scattering phase shifts.  The ratio of the nonlocal wave function $\psi _{NL}(r)
$ to the local equivalent one is, by Eq.\ (\ref{trans}), given by the
expression 
\begin{equation}
F(r)=\frac{1}{\sqrt{1-\rho (r)}}.  \label{Frrho}
\end{equation}%
This ratio is energy dependent, but has no variation with the orbital
angular momentum. The energy dependence of $F(r)^{2}$ arises from the
variation of $\rho _{0}$, $r_{\rho }$ and $a_{\rho }$ with the incident
energy. Table \ref{Table3} shows the maximum values of $F(r)^{2}$ calculated
as a function of the incident energy. The enhancement or suppression effects
of the nonlocal wave function depend on the value of $\rho(r)$ as given by
Eq.\ (\ref{rho}). Table \ref{Table3} shows that $\psi _{NL}(r)$ is enhanced
for all incident energies since $0 \leq \rho(r) < 1.0$.

\begin{table}[ht]
\begin{tabular}{lcc}
\hline \hline
&  &  \\[0.2ex] 
$E_{\mathrm{lab}}$ (MeV) &  & $F(r)^2$ \\ 
&  &  \\ \hline
&  &  \\[0.1ex] 
$12$ &  & 1.82 \\ 
$14$ &  & 1.73 \\ 
$16$ &  & 1.57 \\ 
$18$ &  & 2.25 \\ 
$20$ &  & 2.00 \\ \hline \hline
\end{tabular}%
\caption{Perey factor $F(r)^2$ as defined in Eq.\ (\protect \ref{Frrho}). The
table displays the maximum values of $F(r)^2$ as a function of the incident
energy.}
\label{Table3}
\end{table}

\section{Conclusions}

\begin{figure}[th]
\includegraphics[height=160mm, width=86mm]{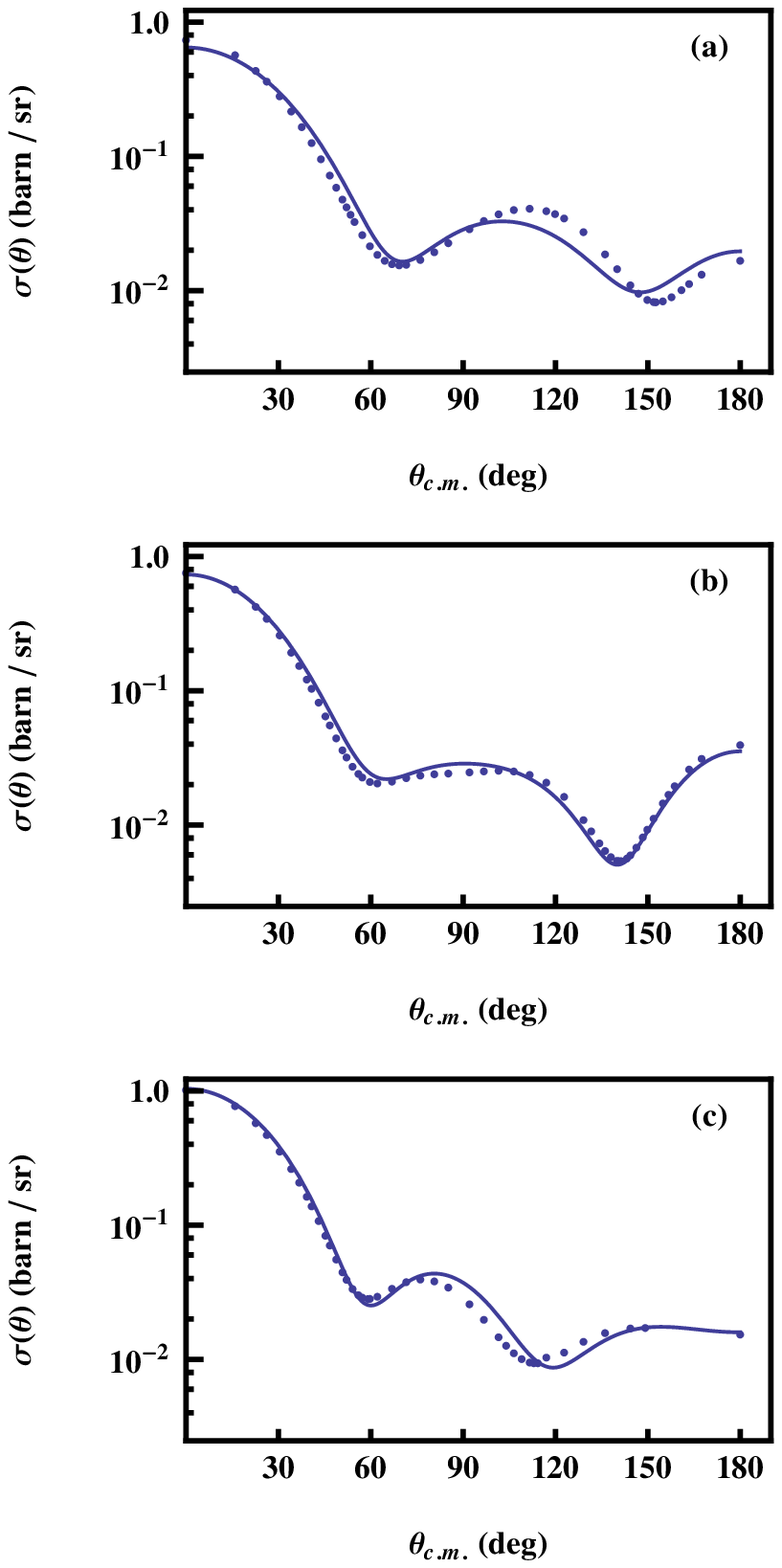}
\caption{The conventional optical potential fits for $N-^{12}C$ elastic
scattering at 12 (a), 14 (b) and 16 (c) all in units of MeV. The model
parameters are given in Table \protect \ref{Table1}. The data is obtained
from reference \protect \cite{ENDF}. }
\label{AngularFits1}
\end{figure}

In a recent work we introduced a velocity-dependent term into the 
phenomenological, conventional optical potential. This resulted in
excellent fits for the $N-^{12}C$ elastic angular distributions even at
large back scattering angles \cite{MJ-GR-PRC}, which is a region known to be
associated with nonlocalities \cite{Rawitscher-L-69,Clark}. In the present
study it has been shown that the velocity-dependent term simulates a source
of nonlocality by showing that inside the target nucleus the nonlocal wave
function $\psi _{NL}(r)$ is enhanced or suppressed compared to the
corresponding local one $\psi _{L}(r)$ obtained in the absence of the
velocity-dependent term. This is in accordance with the Perey effect \cite%
{Perey}. In Table \ref{Table1} we show the fit parameters for the
equivalent local potential $V(r)$ given in Eq.\ (\ref{Vr}) for the $N-^{12}C$ elastic scattering
process. Clearly, the parameters are energy sensitive, and the strength of
the local central potential $V_{0}$ shows the largest variation with
incident energy. This indicates the presence of nonlocal effects in the
scattering process. In this work and in Ref.\  \cite{MJ-GR-PRC} we proposed
that the gradient term simulates a nonlocality, which is due to a change in
mass of the incident nucleon when moving inside the nuclear matter of the
target nucleus. The corresponding fit parameters for the nonlocal,
velocity-dependent potential $V(r,p)$ (given in Tables I and II of reference 
\cite{MJ-GR-PRC}) show less variation with incident energy. This clearly
indicates that there are other sources of nonlocalities like, for example,
Pauli exchange and knock out processes in addition to the detailed structure
of the target nucleus that still need to be accounted for. The local
potential angular distribution fits are given in Figure \ref{AngularFits1}, 
which show that the details of the angular distributions are less
well described compared to the corresponding ones obtained in Ref. 
\cite{MJ-GR-PRC}.

\begin{figure}[ht]
\includegraphics[height=120mm, width=86mm]{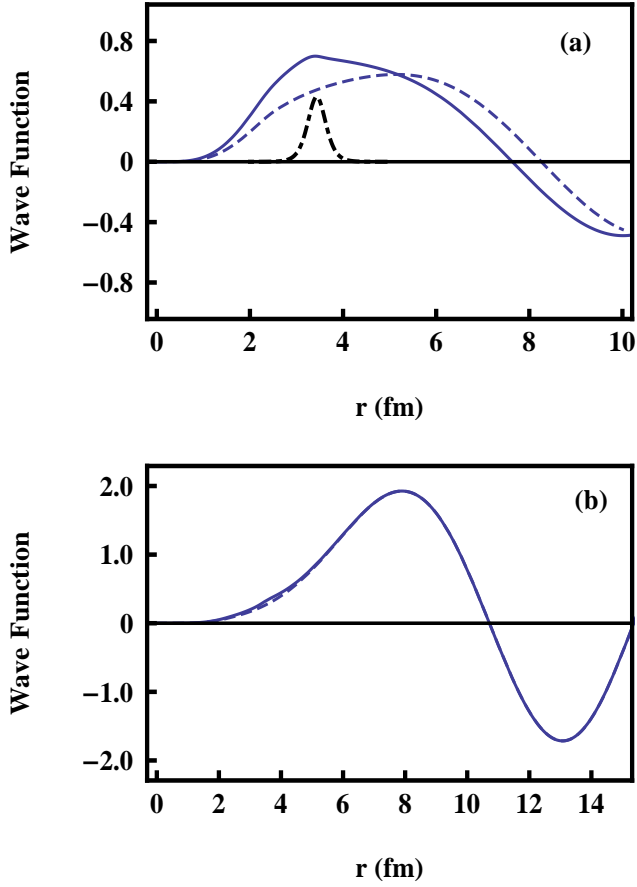}
\caption{ The real parts of the local (dotted) and nonlocal (solid) wave
functions corresponding to the following angular momentum quantum numbers:
(a) $l=3$, and (b) $l=4$. In (a) the dash-dotted peak corresponds to $%
\protect \rho(r)$ defined in Eq.\ (\protect \ref{rho}). The peaks of $\protect%
\rho(r)$ and the nonlocal wave function almost coincide, which explains the
distorted peak of the latter. The local and nonlocal wave functions correspond to the local and nonlocal potentials given by Eqs.\ (\ref{vdp}) and (\ref{Vr}). Both figures correspond to $E=14$ MeV incident
neutron energy.  }
\label{wfncs2}
\end{figure}

We have also shown that the enhancement or  suppression effects are
angular momentum dependent. In Table \ref{Table2} we show the Perey factor $%
F(r)^{2}$ as a function of the incident energy as well as the orbital
angular momentum quantum number $l$. The angular momentum dependence of $%
F(r)^{2}$ is suggestive of the inclusion of a parity dependent term $(-1)^{l}
$ in the conventional optical potential \cite{Rawitscher-L-69,RG_94}. Since
the gradient term simulating the nonlocality is purely real only the real
part of $\psi _{NL}(r)$ is affected and the imaginary part is almost
identical with the corresponding one for the local wave function. Figs.\  \ref%
{wfncs1} and \ref{wfncs2} show the real parts of local (dotted) and nonlocal
(solid) wave functions for different angular momentum quantum numbers $l$
all at the same incident energy of 14 MeV. Clearly, $F(r)^{2}$ is a function
of orbital angular momentum.

Finally, we considered two methods to determine the local wave function $%
\psi _{L}(r)$ corresponding to the nonlocal one, $\psi _{NL}(r)$, that was
obtained in Ref. \cite{MJ-GR-PRC}. The first corresponds to obtaining the
best fit parameters for an equivalent local  potential such that $%
\psi _{L}(r)$ and $\psi _{NL}(r)$ are phase equivalent. The other obtains a
trivially equivalent local wave function, which starts by fitting the data
in the presence of the velocity-dependent term and then mathematically
transforming the $d\  \psi _{NL}(r)/dr$ away by a renormalization of the wave
function as given by Eq.\ (\ref{trans}). In this case the nonlocal $\psi
_{NL}(r)$ and trivially local $\chi _{L}(r)$ wave functions share the same
fit parameters and, by Eq.\ (\ref{Frrho}), $F(r)^{2}$ has energy dependence
only as shown in Table \ref{Table3}. Since the values of $\rho (r)$, which
is given by Eq.\ (\ref{rho}), range from zero to less than unity $\psi
_{NL}(r)$ is enhanced for all energies.

In conclusion, the non local nature of the derivative term is justified here
in terms of the properties of the corresponding Perey Damping Factor, which
differs from unity by 20 to 30\%. Furthermore, in future works we shall test
the applicability of the introduced velocity-dependent potential to other
systems like $p-^{16}O$ and $n-^{40}Ca,$ in order to determine whether%
the surface term for $\rho (r)$ as defined in Eq.\ (\ref{rho}) is
still valid for heavier nuclei, or whether it should be
replaced by a volume term that is proportional to the nuclear density
itself, rather than its gradient.

\end{document}